\documentclass[prd,superscriptaddress,onecolumn,showpacs,11pt,onehalfspace,amsmath,preprintnumbers,showkeys]{revtex4}
\usepackage{wrapfig,rotating}
\usepackage{graphicx,epsfig,color}


\usepackage{graphicx}
\usepackage{dcolumn}
\usepackage{bm}

\def\beq{\begin{equation}}
\def\eeq{\end{equation}}
\def\beeq{\begin{eqnarray}}
\def\eeeq{\end{eqnarray}}

\newcommand \Pomeron {I\!\!P}
\def\2GPD{$_2\mbox{GPD}$}

\def\12{$1\otimes 2$}
\def\22{$2 \otimes 2$}

\def\tr{\rm tr}

\begin{document}

\title{Shadowing
in multiparton
 proton - deuteron collisions}
\author{ B.\ Blok}
\affiliation{Department of Physics, Technion---Israel Institute of
Technology, 32000 Haifa, Israel} \email{blok@physics.technion.ac.il}
\author{ M.\ Strikman}
\affiliation{Physics Department, Penn State University, University
Park, PA, USA} \email{strikman@phys.psu.edu}

\begin{abstract}
We study the screening  effect for the multiparton interactions (MPI) for proton--deuteron collisions in the kinematics where one parton
belonging to  the deuteron has small $x_1$ so the leading twist shadowing is present while the second parton ($x_2$) is involved in the interaction
in the kinematics where shadowing effects are small.
We find that   the ratio of the shadowing and the  impulse approximation terms is approximately factor  of two larger for MPI than for the single parton distributions. We also calculate  the
double parton antishadowing  (DPA) contribution to the cross section  due to the independent interactions of the partons of
 the projectile proton
 with two nucleons of the deuteron and find that shadowing leads to a strong reduction of
 the DPA effect.
 For example, for the
resolution scale $Q_1^2 \sim 4$~$\mbox{GeV}^2$  of the  interaction with parton $x_1$   we find that shadowing reduces the
DPA
effect by $\sim$ 30\%. It is argued that in the discussed kinematics  the contribution of interference diagrams, which correspond to the interchange of
 partons
 between the proton and neutron, constitutes only a small correction to the  shadowing contributions.
\end{abstract}

\maketitle
\setcounter{page}{1}

\section{Introduction}
\par Recently there was a renewed interest
in the theoretical studies of the multiparton interactions  (MPI) in which at least two partons of one of the
 colliding particles are involved in the proton - nucleus collisions\cite{ST,BSW,DS1,DS2,GT1,GT2,GT3}.
 To large extent this
 is due
  to the first experimental studies of
   $pA$ 
  collisions at the LHC \cite{Alice,Alice1,Atlas,CMS}. It was suggested in \cite{ST,BSW,DS1,DS2} that    MPI  would be easier to observe experimentally in $pA$ collisions than in $pp$ collisions since they are parametrically enhanced in the $pA$ case
  by a factor $A^{1/3}$ \cite{ST}. General formulae for this cross section
  were
  derived in  \cite{BSW} within  perturbative QCD (pQCD) in the impulse approximation (that is neglecting deviations of the nuclear parton distribution functions (pdf) from the additive sum of the nucleon pdfs). The analysis demonstrated connection of the pQCD treatment
  with
   the parton model calculation of \cite{ST} for the large $A$ limit and uncorrelated nucleon distribution in the nucleus.
\par   The calculation of \cite{BSW} employed   the formalism developed
in Refs. \cite{BDFS1,BDFS2,BDFS3,BDFS4}, which is based on the use of the  generalized double parton distributions
in momentum space introduced in Ref. \cite{BDFS1}. The calculation was done explicitly in the impulse approximation.
\par It was argued in Refs. \cite{GT1,GT2,GT3}, that  the impulse approximation is not
a complete answer and one must include also the so called interference diagrams, although no explicit estimates
of their
relative strength was
performed.
In Ref.  \cite{BSW} the arguments were presented
that interference diagrams become important for small x due to the leading twist (LT)  nuclear shadowing phenomenon.

\par The
main
aim of the paper is to  calculate explicitly the interference corrections to the impulse
approximation due to the nuclear shadowing for the case of proton - deuteron scattering
based on
the theory of the leading twist shadowing phenomena (for a recent review see
   \cite{Guzey}) which successfully predicted gluon shadowing for the coherent photoproduction of $J/\psi$ recently observed at the  LHC \cite{psi}.
We will focus on the limit when one of partons in the deuteron has small enough $x$, so that nuclear shadowing is present for the deuteron pdf while the second parton is probed in the kinematics where shadowing effects are absent.
We will demonstrate  that in this limit nuclear shadowing induced interference is present already on the level of diagrams where one of the nucleons is active in the amplitude and two in the conjugated amplitude (or vise versa),
and that it
has the same magnitude as the enhancement of MPI due to the interaction with two nucleons in the impulse approximation.
 In our analysis we will neglect a small effect of antishadowing in the deuteron pdfs at $x \sim 0.1$ which is present due to the momentum sum rule, see discussion in \cite{Guzey}.
 We also consider the interference for the case when just one parton of proton is interchanged with one parton of  neutron and argue that this interference effect is much smaller than the leading twist shadowing interference.

 While
the actual experiments are done with the heavy nuclei, we believe that the deuteron case
provides
 a
simple
"laboratory" for the studying possible mechanisms of shadowing in four jet
production
processes.
 In the case of heavy nuclei combinatorics  of the shadowing diagrams is
 much more complicated.  It will be considered  elsewhere.

  The shadowing in the multijet production  differs significantly from the LT shadowing for nuclear pdfs
  since the two
partons belonging to the projectile proton are typically located in a very small transverse area of the radius $\sim 0.5 fm$. As a result they scatter off two different  but very close in the  impact parameter space nucleons that may be rather strongly correlated. This is especially true for
 the case of scattering off the deuteron which is  a highly correlated system.
Hence the  analysis presented here can serve as  the stepping stone to a discussion of similar effects for
MPI with heavy nuclei.

In the current experimental studies
 one usually starts with a trigger on a hard process of large virtuality  - say dijet with $p_t$'s larger than 50 $\div$ 100  GeV  and one next looks for  a second hard subprocess in the underlying event.
Since  the LT nuclear shadowing  for
$p_t \ge $ 100 GeV/c
is  very small we will focus here on consideration of the MPI in which one of the subprocesses has large  enough x or large virtuality so that
the leading twist nuclear shadowing can be neglected in this case.
\par
The paper is organized as  following.
In section 2 we apply  the general expressions relating
double hard four jet cross section for the collision of hadrons $A$ and $B$
in terms of $_2$GPDs (Eq.~\ref{1})
 to obtain  a compact  expression for the
double  parton antishadowing  contribution (DPA)
taking into account the finite transverse size of the gluon GPD in the nucleon.
In section 3 we summarize
first
the theory of the LT shadowing for the deuteron pdfs and
 next use it to
calculate the shadowing correction to the MPI rate
for  the case when x of one of the partons of the  deuteron participating in collision is large and another is small.
We demonstrate that the shadowing in the case of MPI is a factor of two stronger than in the case of the deuteron pdfs. At the same time an additional  contribution to MPI due to the pQCD evolution induced correlations in the proton wave function reduces this enhancement.
In section 4 we present the numerical results. We  find  that shadowing effect is smaller but of the same magnitude as DPA for modest virtualities  ($Q^2\sim 4 \mbox{GeV}^2$).
We
show explicitly that the double parton shadowing is negligible when both of the partons have large $x$, confirming the results of Ref.~\cite{BSW}.
In section 5 we  estimate the contribution of the interference diagrams   corresponding to the situation when a parton "1" ("2") in the amplitude  belongs to the proton (neutron) and in the conjugated amplitude to the neutron (proton). We argue  that these contributions are small compared to shadowing mechanisms.
 Our conclusions are presented in section 6.
In the Appendix we consider correspondence of the Glauber series for the inelastic $pA$ scattering and combinatorics of MPI.

\section{Impulse approximation for the proton - deuteron  scattering}
\subsection{Leading term}
\par Let us first consider the case when  both partons of the nucleus involved in the interaction belong to the same nucleon -- the impulse approximation.
(Fig.~1).
 \begin{widetext}
\begin{center}
\begin{figure}[h]
 \includegraphics[height=4cm]{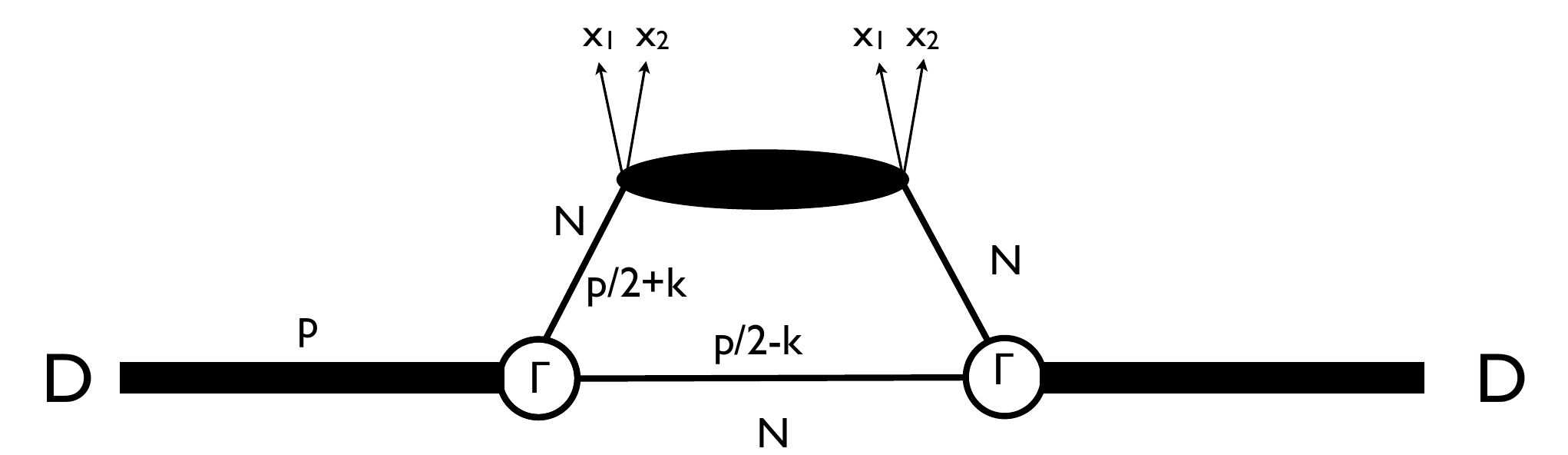}
\caption{\label{Fig1}Impulse approximation.}
\end{figure}
\end{center}
\end{widetext}

 This is the dominant contribution in the deuteron case, though
 it becomes subleading for heavy nuclei \cite{ST,BSW}.
 The corresponding cross section is, obviously,  twice the cross section of the  MPI $pp$ scattering (we neglect here difference of  the quark distributions in  proton and neutron). It
is given by
\beq  \sigma_{\rm imp 4}(pD) = 2 \sigma_{\rm imp 4}(pN).
\eeq
So introducing so called $\sigma_{eff}(pD)$ we can write
\beq
\frac{1}{\sigma_{eff \,pA}} =\frac{\sigma_{\rm imp 4}}{\sigma_1\sigma_2}=2\int\frac{d^2\Delta_t}{(2\pi)^2} F_{2g}(\Delta^2,x_{1})F_{2g}(\Delta^2,x_{2})F_{2g}(\Delta^2,x_{1p})F_{2g}(\Delta^2,x_{2p})(1+N),
\label{2}
\eeq
where $\sigma_1,\sigma_2$ are the elementary cross sections of production of jets in the parton - parton interaction;
the factor $F_{2g}$ is the two gluon form factor of nucleon \cite{FSW}.
The factor  $1+N$ parameterizes  the enhancement  of the observed cross section as compared to  the calculation in the mean field approximation.

A significant positive contribution to $N$ originates from the pQCD evolution
induced parton - parton correlations - the \12  processes
\cite{BDFS1,BDFS2,BDFS3,BDFS4} which enhance the  cross section as compared to the one calculated assuming dominance of the collisions of two independent pairs of partons - the \22 processes. Our numerical studies found
$1+N\sim 2.2$
for $pp$ scattering in quasi symmetric kinematics
 which is  consistent with the
 LHC
 data for
$x\sim 0.001\div 0.01$.
In the kinematics we consider here - one large $p_t$ pair of $p_t \sim 30$ GeV/c
 jets and another pair with  moderate $p_t$'s   of the order $2, 3, 10$ GeV/c the mechanism of \cite{BDFS1,BDFS2,BDFS3} leads to an expectation of $N\sim 0.3, 0.6, 1.0$ respectively. Note that
these values of $N$ are
slightly  larger than  the corresponding values in $pp$ collisions at the LHC for the same hard
transverse scales, since the c.m. energy in $pA$ collisions is smaller
($\sqrt{s}=5~$ TeV) and corresponding $x$
are larger by a factor 1.3
than in
$pp$ collisions for $\sqrt{s}= 8~$TeV.

\subsection{Antishadowing contribution.}
 The second contribution, which  becomes dominant in the case of scattering off heavy nuclei,
 results from the process in which
  two partons from a incoming proton  interact with two different
nucleons of the deuteron.  Corresponding diagram
 is depicted in Fig.~2.

 \begin{widetext}
\begin{center}
\begin{figure}[h]
 \includegraphics[height=5cm]{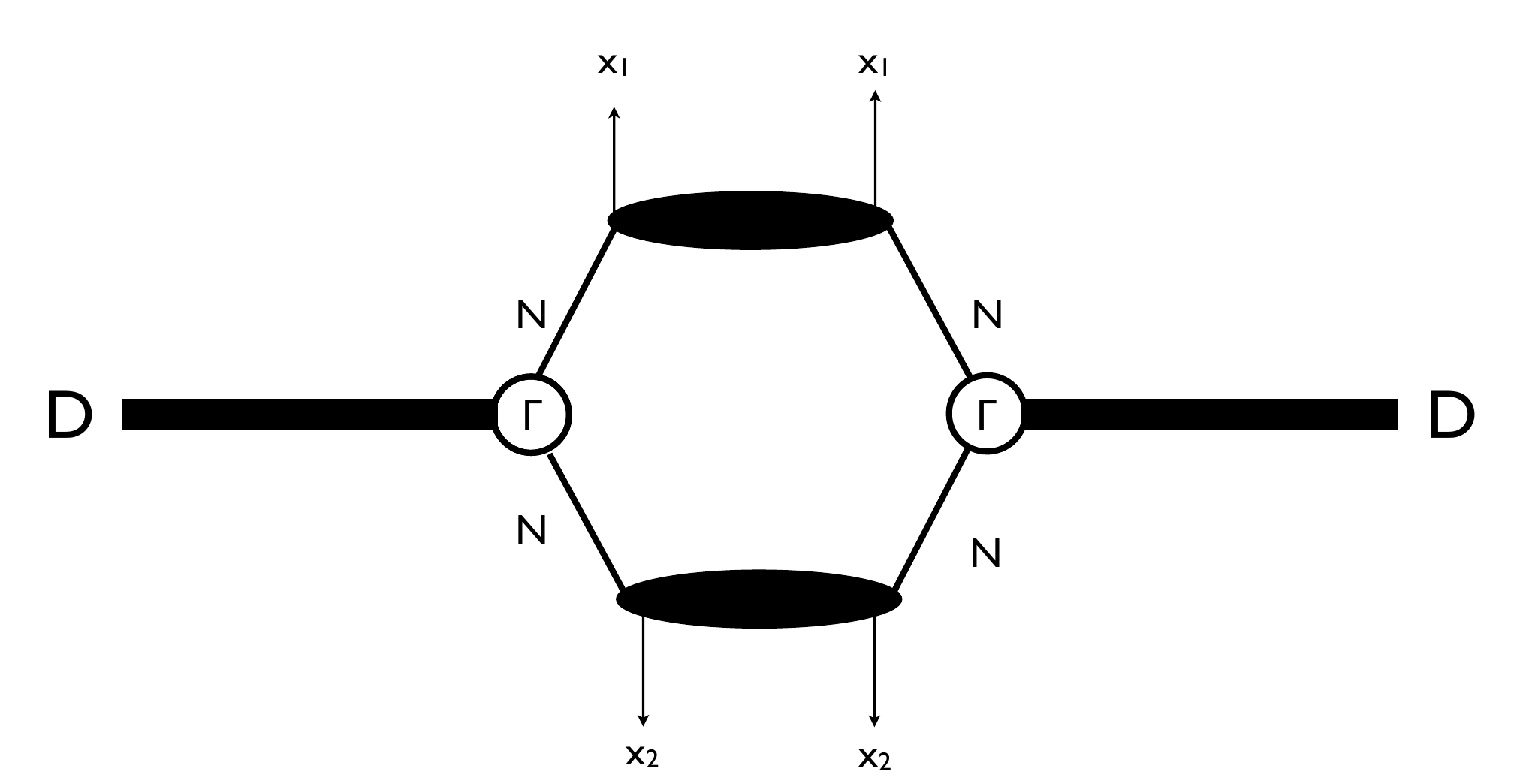}
\caption{\label{Fig2}Double parton antishadowing correction.}
\end{figure}
\end{center}
\end{widetext}

 It can be calculated using the general expression  relating
double hard four jet cross section for the collision of hadrons $A$ and $B$
in terms of $_2$GPDs
\beq
\frac{d\sigma^{AB}_{4jet}}{d\hat t_1d\hat t_2}= \int\frac{ d^2\vec\Delta}{(2\pi)^2}
\frac{d\hat{\sigma}_1(x_1',x_1)}{d\hat t_1}\frac{d\hat{\sigma}_2(x_2',x_2)}{d\hat t_2}
 \, \, _2G_{A}(x_1',x_2',\vec\Delta)\, _2G_{B}(x_1,x_2,\vec\Delta)
  \, ,
 \label{1}
 \eeq
where in our case $G_A, G_B$ are the
$2$ parton GPDs of the nucleon and the deuteron\cite{BDFS1}.
Here $x_1'=x_{1p},x_2'=x_{2p}$ are the light-cone fractions for the  partons of the   projectile nucleon, and $x_1, x_2$ are the light-cone fractions for   the target nucleon/nucleons.
It was demonstrated in \cite{BSW} that
this contribution can be written through the  two - body nuclear form factor. In the case of scattering off the deuteron
(diagram of Fig.~2)
this form factor  is easily
 calculated and expressed through the deuteron form factor (since in this case there is a simple relation between two -  body and single -  body form factors).
Indeed,
the contribution of the
corresponding diagram is given by (cf. Fig.~2 and Eqs. 19 - 21 in \cite{BSW}).
\begin{eqnarray}
\frac{\sigma_{\rm DPA}}{\sigma_1\sigma_2}&=&2\times \int \frac{d^4\Delta }{(2\pi))^4}F_{2g}(\Delta_t,x_1)F_{2g}(\Delta_t,x_2)F_{2g}(\Delta_t,x_{1p})F_{2g}(\Delta_t,x_{2p})\nonumber\\[10pt]
&\times&\int\frac{d^4k}{(2\pi)^4} \frac{\Gamma (p/2 +k, p/2 -k)\Gamma (p/2 +k-\Delta, p/2 -k+\Delta )}{((p/2+k)^2-m^2)((p/2+k-\Delta )^2-m^2)((p/2-k)^2-m^2)((p/2-k+\Delta )^2-m^2)}.\nonumber\\[10pt]
\label{ad1}
\end{eqnarray}
The factors $\Gamma$ are the two
vertex functions  depicted in Fig.~2.
We can now integrate in a standard way over $k^0,\Delta^0$, and use the fact that the corresponding denominators are dominated by nonrelativistic
kinematics: $k^0,\Delta^0\sim \vec k^2/M$, and the longitudinal transfer $\Delta_z=0$. After performing the integration
we immediately obtain:
\beq
\frac{\sigma_{\rm DPA}}{\sigma_1\sigma_2}=2\times \int \frac{d^2\Delta_t }{(2\pi))^2}F_{2g}(\Delta_t,x_1)F_{2g}(\Delta_t,x_2))F_{2g}(\Delta_t,x_{1p})F_{2g}(\Delta_t,x_{2p})S(\vec\Delta^2).
\label{11}
\eeq
 We define here the deuteron form factor as (see e.g. \cite{Guzey}):
\beq
S(\vec \Delta^2)=\int \frac{d^3k}{(2\pi)^3 8M}\frac{\Gamma (\vec k^2)\Gamma ((\vec k-\vec
\Delta )^2)}{(A^2+\vec k^2)(A^2+(\vec k-\vec \Delta )^2)},
\label{11b}
\eeq
where the $\Gamma$ is the  deuteron  to
 two nucleons vertex, and
\beq
A^2=m^2-M^2/4.
\label{11a}
\eeq
Here $M$  is the deuteron mass, $m$ is the nucleon mass, and the momenta of nucleons in the  deuteron are $\vec p/2+\vec k$, and $\vec p/2-\vec k$.
Here we used the fact than the deuteron is a nonrelativistic system,
so the form factors $\Gamma$ depend
only on the differences
of
the spacial
components of the nucleon momenta.
Using the relation  between the vertex functions and wave functions of the deuteron we can rewrite the latter expression in terms
of the deuteron nonrelativistic wave functions as
\begin{eqnarray}
S\left(\Delta^2\right)=\int d^3 \vec{p} &&\Bigg[u(\vec{p})u(\vec{p}+\vec{\Delta}) \nonumber \\
&&+ w(\vec{p})w(\vec{p}+\vec{\Delta}) \left(\frac{3}{2}\frac{(\vec{p}\cdot (\vec{p}+\vec{\Delta}))^2}{p^2 (p+\vec{\Delta})^2}-\frac{1}{2}\right) \Bigg]  \,,
\label{eq:rho_D}
\end{eqnarray}
where $u$ and $w$ are the $S$-wave and $D$-wave components of the deuteron wave function respectively (here in difference from Eq.~\ref{11b} we give the expression for the spin-1 deuteron).

Note that Eqs.~\ref{11},~\ref{11b} accurately take into account the finite transverse size of the nucleon GPDs which is numerically rather important (see section 4).

At the same time
we neglected in this calculation the nucleon Fermi motion effect which shifts
the x-argument of the bound nucleon pdfs.
The reason is that  this effects is a very small correction which enters only on the level of the terms $\propto \vec k^2/m^2$ which are very small for the deuteron,
 cf. discussion in \cite{BSW}.

Finally, let us mention that we must multiply this expression by  $1+N_L$, where $N_L$ is the enhancement of 4 jet cross section relative
to mean field approximation in the given kinematics due to parton correlations. In our kinematics this number is very small. Indeed,
in difference from the case of  $pp$ collisions the $\Delta$ dependence of the nucleus and nucleon factors in the corresponding equation is very different.
As a result one does not have  in this case an enhancement factor of $\sim 2$ from \12 which is present in the $pp$ case.
In addition, the transverse integral is dominated by the same deuteron form factor both in $1\otimes 2$ and $2\otimes 2$ contributions,
leading to $N_L\sim N/5\le 0.1$ (see section 4) for $Q^2\le 10$ GeV$^2$, and reaching $20\%$ for $Q^2\sim 10$ GeV$^2$.

\section{Single shadowing: one to two processes.}

\subsection{Leading twist shadowing for the deuteron pdfs}
Before  discussing the shadowing for MPI in the deuteron it is worth reminding the picture of the LT shadowing for the case of the deuteron pdfs.
It was demonstrated in \cite{FS99} that the shadowing correction to the deuteron pdf  can be expressed in  the  model independent way through the diffractive nucleon pdfs.
In the reference frame where deuteron is fast,  the process can be pictured as the hard interaction in $\left| in\right>$-state with a small $x$ parton in which the nucleon in the final state carries most of its initial momentum fraction -- $(1 - x_{\Pomeron})$, while
 in the final
state the diffractive system  which carries the light-cone fraction $x_{\Pomeron}$ combines with the second nucleon  into a nucleon with momentum fraction $1 + x_{\Pomeron}$, see Fig.~3.
\begin{widetext}
\begin{center}
\begin{figure}
 \includegraphics[height=4.5cm]{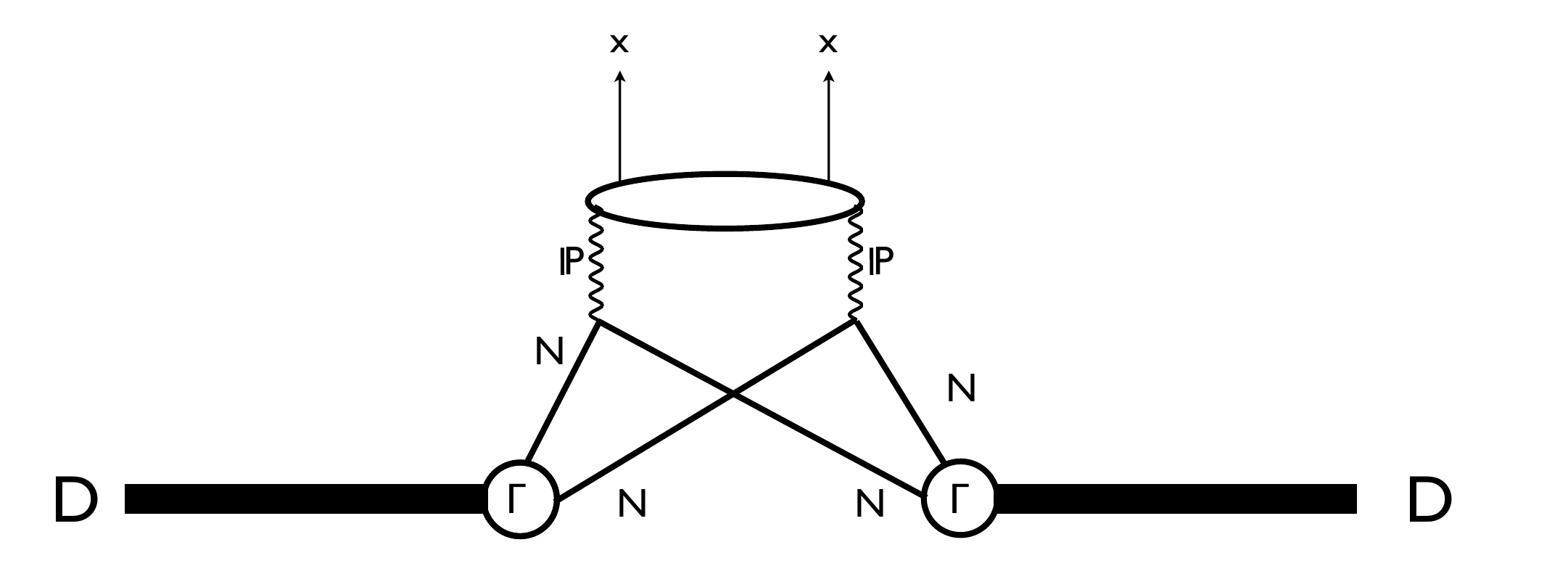}
\caption{\label{Fig3} Shadowing in DIS off the  deuteron.}
\end{figure}
\end{center}
 \end{widetext}

As a result one finds for the shadowing correction  (see Eq. 98 and  Fig.~28   in Ref. \cite{Guzey})
\beq
\Delta f_D(x,Q^2) =  2f_N(x,Q^2)-f_D(x,Q^2),
\eeq

\beq
\Delta f_D=
2\int \frac{d^2 q_t dx_{\Pomeron}}{(2\pi)^3}S(\vec{q}~^2))F^{D(4)}(\beta ,Q^2,x_{\Pomeron},q_t),
\label{23}
\eeq
where $\beta=x/x_{\Pomeron}$ and $F^{D(4)}(\beta ,Q^2,x_{\Pomeron},q_t)$ is the diffractive pdf. It is easy to see that the shadowing originates from configurations where two nucleons are roughly behind each other. For these configurations shadowing is large as long as the effective cross section of the rescattering:
\beq
\sigma_2\approx 16\pi {\int_x^{0.1}dx_{\Pomeron} \beta  F_j^{D(4)}(\beta, Q^2,x_{\Pomeron},t_{min})\over xf_{j/N}(x,Q^2)},
\eeq
is comparable to the pion - nucleon cross section which is
the case  for the gluon channel for $x \le 10^{-3}$, $Q^2\le \mbox{10 GeV}^2$.

The leading twist shadowing theory  \cite{FS99,Guzey} predicted  reduction of the gluon pdfs  in the gluon channel for $x\sim 10^{3}, Q^2 \sim \mbox{few GeV}^2, A= 200 $
by a factor $0. 5 \sim 0.6$ which agrees well with the $J/\psi $ coherent photoproduction data \cite{psi}.  It is worth emphasizing that the expressions for shadowing contribution to the the deuteron pdfs can be derived   both  using pretty cumbersome approach of the original paper of Gribov \cite{Gribov:1968jf} or using Abramovski, Gribov Kancheli (AGK) cutting rules \cite{AGK} in combination with the  the QCD factorization theorems for diffraction scattering and  for inclusive scattering
\cite{Guzey}.  The dominance of the soft Pomeron dynamics for the hard diffraction is now confirmed by the HERA data -- $\alpha_{\Pomeron}$ for hard diffraction is the same as for soft processes \cite{HERA,Guzey}.  So we are  applying AGK rules effectively  for the soft dynamics where it appears to be well justified
 \footnote{In pQCD color effects complicate application of the AGK cutting rules for the inelastic intermediate final states. However the AGK  relation between total cross section and diffractive cut appears to hold  (A.~Mueller, private communication).}

\subsection{Single shadowing for MPI}

The DPA
 contribution which we considered above corresponds
to collisions where two nucleons of the deuteron are located at  small relative transverse distance of the order of the nucleon
transverse gluon size - $\sim 0. 5 fm$. For  such two nucleon configuration  LT nuclear shadowing is large since  the effective cross section of the rescattering interaction is large. Hence it may  strongly reduce the DPA
effect.
The shadowing term corresponds to the diagrams which are an  analog of the LT shadowing diagrams for the deuteron pdf with an extra blob corresponding to the non screened second interaction (Fig.~4).
\begin{widetext}
\begin{center}
\begin{figure}
 \includegraphics[height=5.5cm]{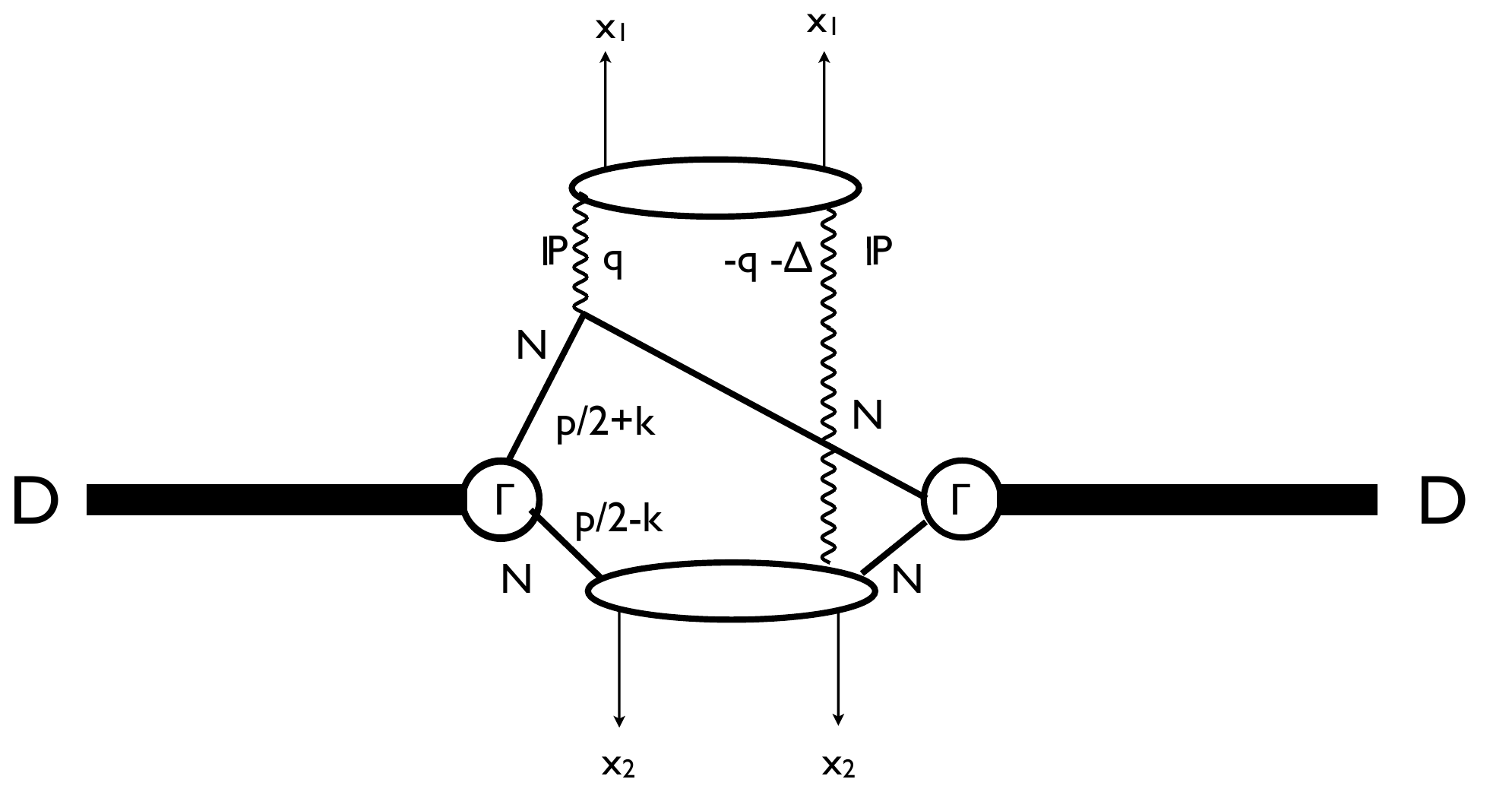}
\caption{\label{Fig4} Shadowing correction to 4 jet production in pD scattering.}
\end{figure}
\end{center}
 \end{widetext}
The screening contribution requires that the first
nucleon  experiences the diffractive interaction, while the second hard blob is a generic hard nucleon nucleon interaction.
 Similar to the DIS case this diagram gives negative contribution to the cross section.

As usual only the diagrams with elastic $\Pomeron - nucleon - \Pomeron$ vertex contribute,   since we work in conventional two  nucleon approximation for the deuteron when all other components of the deuteron wave function are neglected.

 Hence  the shadowing is described by
  four diagrams one of which is depicted in Fig.~4.
  The combinatorial factor of two arises   since the parton "1" can belong to either of two nucleons. Another  factor of two is due to the possibility to attach the Pomeron line to the first nucleon either in the initial or in the final state.
The shadowing contribution
  can be written as:

 \begin{eqnarray}
\frac{\sigma_{SS}}{\sigma_1\sigma_2}&=&-4\int \frac{d^4qd^4\Delta d^4k}{(2\pi)^{12}}
\frac{F^{D(4)}(\beta,Q_1^2,q^2_t,x_{\Pomeron},\vec \Delta_t)}{G_N(x_1,Q_1^2)}\frac{1}{((p/2+k)^2-m^2)((p/2+k-q+\Delta)^2-m^2)}\nonumber\\[10pt]
 &\times&\frac{F_{2g}(\Delta_t,x_{1p})F_{2g}(\Delta_t,x_{2p})F_{2g}(\Delta_t,x_2)}{((p/2-k)^2-m^2)((p/2-k-\Delta)^2-m^2)((p/2-k-\Delta+q)^2-m^2)}+(1\leftrightarrow 2),\nonumber\\[10pt]
\label{22c}
\end{eqnarray}
with the factor of four reflecting presence of four diagrams.
The Pomeron exchanges carry three-momenta  $\vec q =(\vec q_t,q_z)$ and $\vec q+\vec\Delta $.

We carry the integration over $q_0,k_0,\Delta_0$ in exactly the
same way as in the previous section, where we calculated
the diagram
of Fig.~2,
taking  into account that the vector $\vec\Delta$ is transverse.
 Using Eq.~\ref{11b} for the  deuteron form factor we can rewrite  Eq.~\ref{22c}  as
\begin{eqnarray}
\frac{\sigma_{SS}}{\sigma_1\sigma_2}&=&-4\int \frac{d^2q_td^2\Delta_t dx_{\Pomeron}}{(2\pi)^5} \displaystyle{\frac{F^{D(4)}(\beta,Q_1^2,q^2_t,x_{\Pomeron},\vec \Delta_t)}{G_N(x_1,Q_1^2)}}S((\vec q+\vec\Delta)^2)\nonumber\\[10pt]
&\times&F_{2g}(\Delta_t,x_{1p})F_{2g}(\Delta_t,x_{2p})F_{2g}(\Delta_t,x_2))
+(1\leftrightarrow 2).
\label{22}
\end{eqnarray}

 Overall,
  we can see from the comparison of Eqs. \ref{23} and \ref{22}
 that in the limit when radius of the deuteron is very  large so one could neglect the $q_t$ dependence of all other factors,  the ratio of  shadowing and impulse approximation terms  in the case of the MPI is a factor of two larger than for case of DIS. This reflects the enhancement of the central collisions in the MPI which we mentioned above.
 Note that we use here implicitly the AGK relation between the cross section for the total MPI cross section and for the cross section for the inelastic final state depicted in Fig.~4. In principle, one could first obtain the expression for the small x parton distribution in the impact parameter space as a function of the transverse distance between the nucleons (cf. \cite{Guzey} where GPDs for the nuclei at small x are calculated) and next calculate the $\rho$ distribution of the second parton, ultimately deriving $_2$ GPD for the deuteron and calculating the MPI cross section using $b$ space representation \cite{Paver:1984ux,mufti}.
However, similar to the case of DPA  the expressions in the  momentum
space representation are more compact.

 \section{Numerical estimates.}
\subsection{Antishadowing.}
For numerical estimates it is convenient to approximate
 the deuteron form factor calculated with a realistic deuteron wave functions by  a sum of two exponentials \cite{LS}
\beq
S(\vec \Delta^2)=0.6\exp(-K^2_{1D}\vec \Delta^2)+0.4\exp(-K^2_{2D}\vec \Delta^2),
\label{13}
\eeq
where
\beq
 K^2_{1D}=22.7 \mbox{GeV}^{-2},K^2_{2D}=127 \mbox{GeV}^{-2}.
 \eeq
 The momentum dependence of the two gluon form factor can be extracted \cite{FSW} from   the $J/\psi $ photoproduction data. The exponential fit gives
\beq
F_{2g}(\vec\Delta^2,x)=\exp(-\vec\Delta^2B_N(x))),
\label{14}
\eeq
\noindent where
\beq
B_N\approx 1.43 +0.14Log[x_0/x]\,\,\, {\rm GeV}^{-2}.
\label{15}
\eeq
and  $x_0=0.1$.
For understand better qualitative features of the interplay between the  distance scales related to the deuteron and to the nucleon GPDs  we shall use below a simplified form of
the deuteron form factor
\beq
S(\vec \Delta^2)=\exp(-K^2_{D}\vec \Delta^2),
\label{s1}
\eeq
while in the numerical calculations we will use Eq.~\ref{13}.
(the radius $K^2_{D}$ is related to the electric  radius of the deuteron as
$K^2_{D}= (2/3) R^2_{D\, e.m.}$.)
Performing integration in Eq.~\ref{2} we obtain for the leading term:

 \beq
\frac{\sigma_{\rm imp4}}{(\sigma_1\sigma_2)}=\frac{1}{2\pi}\frac{(1+N)}{K(x_1,x_2,x_{1p},x_{2p})},
\label{ref15a}
\eeq
where
\beq
K(x_1,x_2,x_{1p},x_{2p})=B_N(x_1)+B_N(x_2)+B_N(x_{1p})+B_N(x_{2p}).
\eeq
The function $K$ is determined by the two gluon form factors
of the  nucleon. It is
independent of the  deuteron
wave function.
The answer for the DPA
correction to the cross section is obtained by taking integral over
$\vec{\Delta}$
in Eq.~\ref{11}
using parametrization \ref{s1}:

\beq
\frac{\sigma_{DPA}}{\sigma_1\sigma_2}=\frac{1}{2\pi}
\frac{1}{K^2_{D}+K(x_1,x_2,x_{1p},x_{2p})}.
\label{16}
\eeq
Using parametrization \ref{13}
for the deuteron form factor, we obtain the  DPA
correction of the order $8\%$
when all $x$'s
are $\sim 0.01$, (neglecting $N_L$) and slowly decreasing with
a further
decrease of x's.
This is in very good agreement with a more explicit calculation
 using a  expression \ref{eq:rho_D} for the form factor and the Paris   deuteron wave functions, which  gives $7.3\%$.
 Note here that neglecting
     the nucleon finite size as compared to the deuteron size (putting $B_N$ to zero in Eq.~\ref{16}) would result in an  overestimate of the  discussed contribution to the cross section by 25$\div$ 30\%.

 \subsection{Single shadowing.}
  We now use the simple parametrization for the nucleon diffractive
  pdf
   $F_D$\cite{Guzey},\beq
F^{4(D)}D(\beta ,Q^2,x_{\Pomeron},q_t)=B_D\exp(-B_Dq^2_t) F^{3(D)}(\beta ,Q^2,x_{\Pomeron}),
\eeq
where $\beta=x_1/x_{\Pomeron}$.
In the limit of small x when we can neglect $t_{\min}=-m^2_Nx^2_{\Pomeron}/(1-x_{\Pomeron})$,
 integral over longitudinal and transverse degrees of freedom in Eq.~\ref{23} decouple. In this
 limit
  Eq.~\ref{23}
for the shadowing correction can be rewritten as  (we can neglect $x_{\Pomeron}$ in the argument of the deuteron form factor)
\beq
\Delta G(x,Q^2)=-I(x,Q^2)B_D(\frac{0.6}{K^2_{1D}+B_D}+\frac{0.4}{K^2_{2D}+B_D})= -S\cdot I(x,Q^2)=-0.166 I(x,Q^2),
\label{24}
\eeq
where we defined
\beq
I(x,Q^2)=\int^{0.1}_x dx_{\Pomeron} \beta F_3(\beta,Q^2,x_{\Pomeron})
/8 \pi^2.
\label{63}
\eeq
and $S$ is the
integral over transverse momenta:
\beq
S=B_D(\frac{0.6}{K^2_{1D}+B_D}+\frac{0.4}{K^2_{2D}+B_D})
\label{64}
\eeq
Here $B_D \mbox{=7~GeV}^{-2}$ is  the slope of diffractive structure function of the nucleon based on the  HERA
experimental data which
 indicates that
 $B_D$
  practically does not depend on
$x_{\Pomeron}$ \cite{HERA}.
In this approximation
the function $I(x,Q^2)$ can be easily determined
 from numerical results for $\Delta G(x,Q^2)$ \cite{Guzey}.

\par Now we can use
expression \ref{22} for the single parton shadowing in four jet production to calculate the value of the shadowing effect.
For the exponential parametrization we can write
\beq
F^{4(D)}(\beta,Q^2,x_{\Pomeron},q_t,\Delta_t)=B_D\exp(-q_t^2B_D/2-(q_t+\Delta_t)^2B_D/2)F_3(\beta,Q^2,x_{\Pomeron}).
\label{29}
\eeq
Hence the shadowing correction is
\beq
\frac{\sigma_{SS}}{\sigma_1\sigma_2}=-{4(I(x_1,Q_1^2)U(x_1,x_2,x_{1p},x_{2p})+I(x_2,Q_2^2)U(x_2,x_1,x_{1p},x_{2p}))\over 4\pi}.
\label{28a}
\eeq
Here the longitudinal function $I$ is given by Eq.~\ref{63} and   the transverse integrals $U$ are obtained by using Eq.~\ref{22},
and
explicit Gaussian parametrization for the  form factor.

The ratio $K=\sigma_{SS}/\sigma_{DPA}$ is presented in Fig.~5 as
 a function of
 $x_1$
   and $Q_1^2$
   for the LHC kinematics of  production of two jets with $p_t = Q$ and $4Q^2=x_1x_{1p}s$, $s=2.5\times 10^7$ GeV$^2$. The
 second $x_2=0.1, Q_2^2=1000$ GeV$^2$ being fixed to stick to the kinematics under discussion.
    In Fig.~6 we also present
 the ratio of the shadowing correction for this kinematics and
 the full impulse approximation
 result.

 \begin{widetext}
\begin{center}
\begin{figure}
 \includegraphics[height=5.5cm]{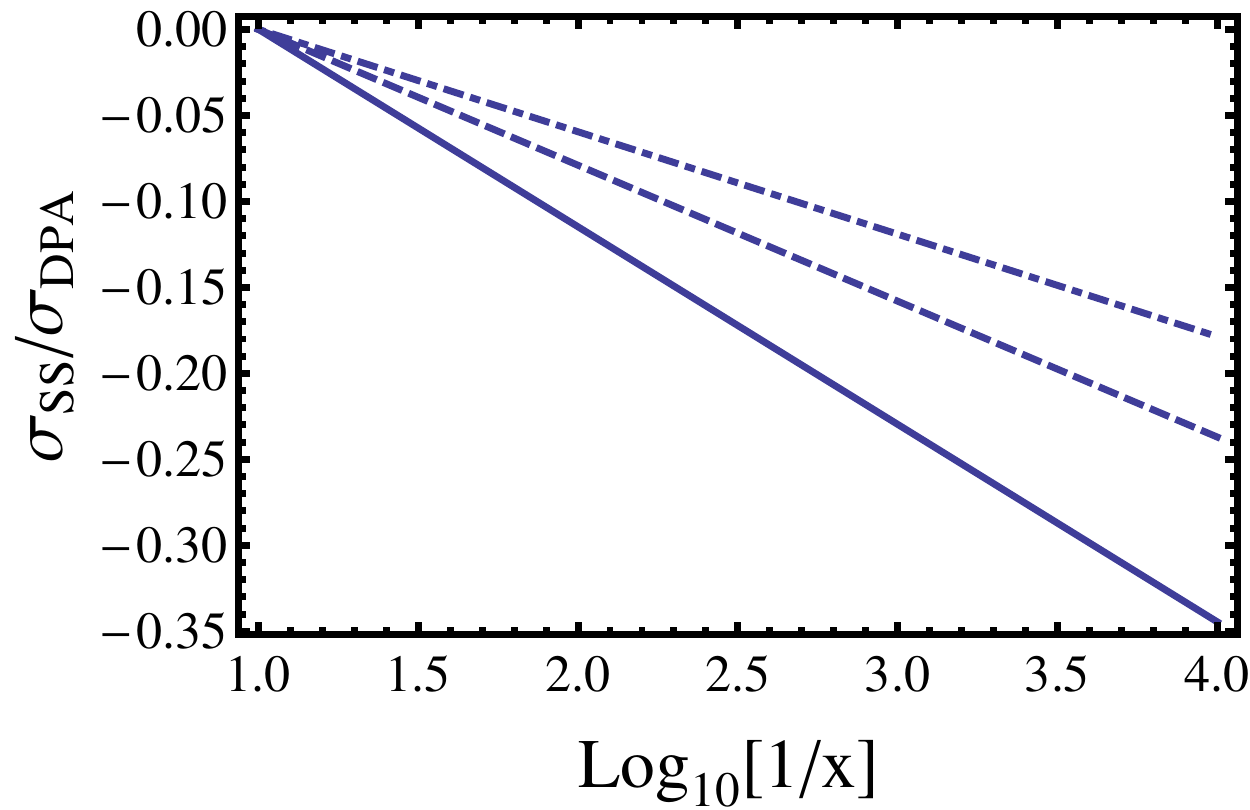}
\caption{\label{Fig5} The ratio $K$ of shadowing and DPA corrections to the four jet production cross section
as a function of $x_1\equiv x$
 for hard scales $Q_1^2=4, 10, 100$ GeV$^2$, $Q_2^2=1000$ GeV$^2$. We put $x_2=0.1$ and $x_{1p}=4Q_1^2/(x_1s), x_{2p}=4Q_2^2/(x_2s)\sim 0.0016$.}
\end{figure}
\end{center}
 \end{widetext}
 \begin{widetext}
\begin{center}
\begin{figure}
 \includegraphics[height=6cm]{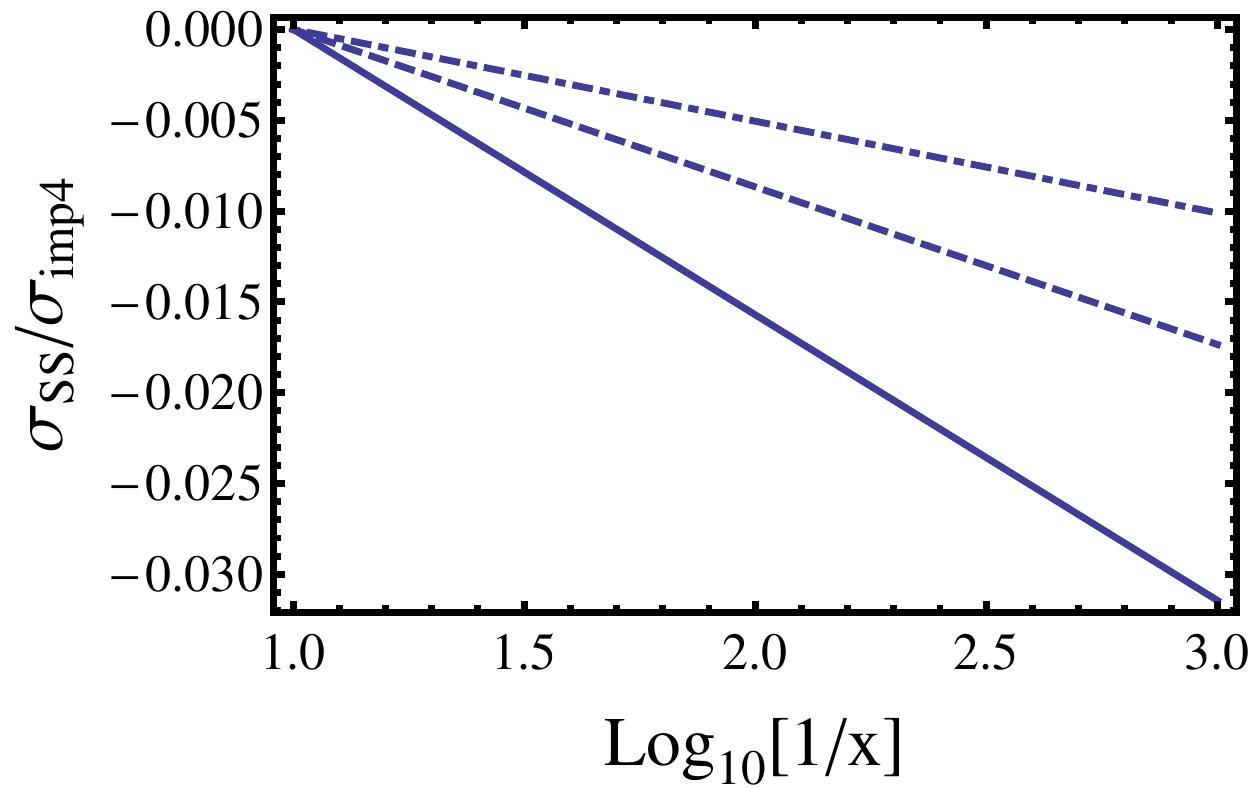}
\caption{\label{Fig6}The ratio of shadowing correction to DPA and full impulse cross section
as a function of
 x for hard scales $Q_1^2=4, 10, 100$ GeV$^2$. $ Q_2^2=1000$ GeV$^2$. We put $x_2=0.1$ and $x_{1p}=4Q_1^2/(x_1s), x_{2p}=4Q_2^2/(x_2s)\sim 0.0016 $.}
\end{figure}
\end{center}
 \end{widetext}

\par  For typical $x_1\sim 0.001,x_2\sim 0.05$ in LHC kinematics we
find
shadowing of order $30\%$ relative to DPA  for low $Q_1^2\sim 4$ GeV$^2$.
We also see from Fig.~5 that the shadowing contribution to the cross section decreases with the increase of the transverse scale.

   Note also that the account for the finite  size of the nucleon
   reduces the  the absolute value of the correction by $\sim 10\%$. The same reduction occurs also for the DPA, so the ratio of shadowing and DPA contributions is practically not sensitive to the finite nucleon radius.

 In the limit of very small $x_1 \le 10^{-3}$  and $x_2 $ large
 one maybe close to the black disk regime  and the LT approximation would break down. Still our calculation indicate that in this limit suppression effect should be large -- $ \sim 0.5$.
relative to DPA.

\par It is instructive to compare
 the shadowing correction to the total differential  cross section  of the  four  jet production in $pD$ collision in the impulse approximation to the shadowing correction to
deuteron structure functions.
The internal over the longitudinal momenta is the same for both corrections and hence
their ratio
is given then by the ratio of transverse integrals
which
is  of the order one.
Indeed,  the ratio of shadowing and impulse contributions can be rewritten as
\beq
\frac{\sigma_{SS}}{\sigma_{\rm imp4}}=\frac{\Delta G_N(x_1,Q_1^2)}{G_N(x_1,Q_1^2}\frac{2}{1+N}\frac{U\cdot K}{S},
\label{101}
\eeq
where we used Eqs. \ref{24},\ref{64}. Thus we see that the shadowing correction
for DPI
is proportional to the shadowing correction to the deuteron gluon PDF,
the proportionality coefficient being the
the product
of the factor  $2/(1+N)$ and the ratio of transverse integrals.
The latter one is always close to one.
 For logarithmic parametrization  of $B_N$ the transverse factor $U\cdot K/S$
does not depend on $x_1$ (only on the hard scales). The factor
2/(1+N) also depends on $x_1$ only  weakly, at least for $x_1\ge 0.001$ and
is
close to one for large $Q_1^2$ while it is of the  order 1.5 at $Q\sim$ few GeV in the chosen kinematics\cite{BDFS3}.

Altogether we see that the $x$-dependence
of the ratio (\ref{101})
 is the same as for  the shadowing correction for the
corresponding deuteron pdf,
 but the absolute value depends on
 the ratio of the
  transverse
integrals (which is  of the order of one) and the value of $N$.
As a result the ratio
is of the order of  $2/(1+N)$.
The factor 2 shows that there is a different combinatorics in MPI in pD scattering and in  the  DIS scattering of the deuteron,
 i.e. one does not obtain the screening
correction simply by substituting the nuclear pdf (that includes shadowing) instead of nucleon pdf in the impulse approximation equations.
\par Finally, let as note that the  ratio $\sigma_{\rm DPA}/\sigma_{\rm imp4}$ of DPA and impulse approximation is x-independent and depends only on hard
scales.
It is
 equal to
\beq
\sigma_{\rm DPA}/\sigma_{\rm imp4}\sim(0.16 \div 0.18)/(1+N),
\eeq
where 0.18 corresponds to the hard scale 4 GeV$^2$ and 0.16 to the 100 GeV$^2$ scale. So  the ratio
slowly decreases with the change of the hard scale,
 mostly due to the change  of N, decreasing from  $\sim 1$
at the hard scale $10$ GeV to $\sim 0.3$ at $2$ GeV, due to the dynamical dependence of N on the scale,
found in \cite{BDFS2,BDFS3}.
\par The $1\otimes 2$ contributions to DPA is small.
Indeed, as it was already mentioned above, there is no factor $2$ that is  present in the $pp$ collisions due to asymmetric kinematics,
Also,
the  integral over $\vec \Delta$, for the $1\otimes 2$ term   in the $ pp$ collisions is proportional to
 $4B_N/2B_N$, enhancing $1\otimes 2$ contributions  by a factor of two  relative to the $2\otimes 2$ contribution. This enhancement however is absent in DPA, where the corresponding ratio
is $(K_D+4B_N)/(K_D+2B_N)\sim 1.1$.  Altogether this results is a strong suppression of the  $1\otimes 2$ contribution in DPA so it can be safely neglected. Similar effect for heavy nuclei was discussed in Ref. \cite{BSW}.

\section{Two nucleon interference. }

It was emphasized  in Refs. \cite{GT1,GT2} that in addition to the impulse approximation mechanism and the double nucleon interaction mechanism considered above there exists a contribution due to the interchange of partons between the nucleons - so that the parton "1" ("2") in $\left|in \right>$ and  $\left<out \right|$ states
belongs to the
  different nucleons.
This is in addition to the interference in nuclear shadowing mechanism which was discussed in section 3.
It was suggested in \cite{GT1,GT2} that such a contribution may give a significant contribution to the cross section,   though no numerical estimates were  presented so far.
 A typical contribution of this kind is depicted in Fig.7 where filled circles represent interactions
with two partons of the projectile. $\alpha_i$ are the light cone fractions carried by proton and neutron and the scale is chosen so that $\alpha_1 + \alpha_2=2$, cf. discussion in \cite{BSW}.

 The interference mechanism is present only for the case when either two (anti)quarks or two gluons are involved in the hard processes and it is absent in the mixed case allowing to avoid completely the interference contribution \cite{GT1,GT2}.
To estimate 
its
 magnitude as compared to the shadowing effects in the kinematics discussed in the paper we need to consider
effects related to the difference of the momentum scales in the deuteron and nucleon as well as the pQCD effects related to the presence of the large scale in the problem. We will consider them in turn.

\begin{widetext}
\begin{center}
\begin{figure}[h]
 \includegraphics[height=8cm]{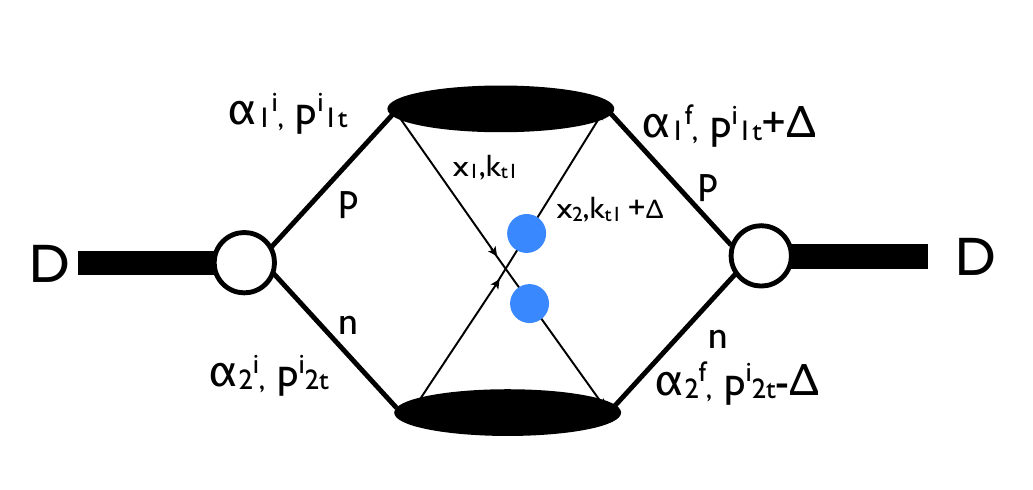}
\caption{\label{Figx}Parton interference  mechanism. The filled circles represent interactions with two partons of the projectile.}
\end{figure}
\end{center}
\end{widetext}

\subsection{Overlap due to the momentum flow}
It was argued in Ref.\cite{BSW} that the interference mechanism is strongly suppressed even in the case of the processes involving (say)  two gluons of the nucleus if $x_1-x_2$ is large enough.  In the case of the deuteron it is possible to elaborate the arguments of  \cite{BSW}. It is straightforward to see that the
integration over the momenta of nucleons in the initial and final states leads to the factor $F_D(\vec{r})$, where
$F_D$ is the deuteron body form factor defined in Eq. \ref{eq:rho_D}, and
$\vec{r}= ((x_1-x_2)m_N, \vec{\Delta}) $ is the 3D momentum transfer to the nucleon of the deuteron calculated in the nonrelativistic limit. Hence in the limit we consider when one $x$ is small and second is far away from the shadowing region there exists a range of $x_1$
\beq
x_1 \geq \sqrt{{3\over 2}} {1\over R_D m_N}  \sim 0.1,
\eeq
where interference is very strongly suppressed by the deuteron form factor independent of the details of the dynamics.

Let us now discuss the interference contribution   for smaller $x_1$  and compare it to  the DPA contribution. First, there are
  generic small factors which are related the dominance of the two nucleon configurations in the  deuteron wave function (accuracy of this approximation is discussed below).

  \par Consider now the dynamical overlap in the final state.
Let us now demonstrate that the overlap integral calculated neglecting color and spin effects is similar to the case of double nucleon interaction.
We consider for simplicity the case when $x_1$ is small and the effect of suppression due to the longitudinal momentum transfer can be neglected.
Also we introduce $\phi^2_N(k_t) $ - the transverse momentum distribution of partons at the low Q - scale which is normalized to one (we do not
 write explicitly its dependence on $x_i$.
The factor $\int d^2\Delta G_N^4(\Delta) S_D(\Delta) $ in the expression for the  DPA contribution is changed to :
\beq R\equiv  \int \Psi_D(p)\Psi_D(p+ \tilde{\Delta}) \phi_N^2(\tilde{\Delta})\phi_N(k_1) \phi_N(k_2) G_N^2(\Delta) d^2  \tilde{\Delta} d^2k_1d^2k_2d^2p.
\label{factor}
\eeq
where $\tilde{\Delta} = - k_1+k_2+\Delta$. The integral over $p$ gives the deuteron form factor $S_D(\tilde{\Delta})$ which  converges on the scale much lower than the parton transverse momentum scale, so in the rest of the integrand we can substitute $\Delta \to k_1-k_2$ and obtain, using Eq.~8:
\beq R =  \int  S_D(\tilde{\Delta}) \phi_N^2(\tilde{\Delta}) d^2 \tilde{\Delta} \int  \phi_N(k_1) \phi_N(k_2) G_N^2(k_1-k_2)  d^2k_1d^2k_2.
\label{factor1}
\eeq
Taking Gaussian transverse momentum distribution for partons in the nucleon:
$\phi_N^2= (1/ \lambda\pi) \exp(-k_t^2/\lambda)$
 with $\lambda= \left<k_t^2\right> \sim 0.25 \mbox{GeV}^2$    we can easily perform integrations and find that
numerically $R$
 is close to the corresponding factor in the expression for the DPA.
    Note here that we considered parton interchange  at a very  low  scale
   $Q^2  \sim 0.25 \mbox{GeV}$. Choosing a more realistic scale $\ge 1 \mbox{GeV}^2$ will lead to a significant
   reduction
   of $R$. The $Q^2$ evolution to the  scale $\sim  p_t^2({\rm jet}) $ leads to an additional suppression which will be discussed below. Hence account of the spacial overlap leads to suppression of interference, so it will be at most of 
    the order of DPA contribution.

\subsection{Suppression of interference in LLA.}
 It was demonstrated in  \cite{BR} that for  the contributions involving the parton interchange are
 suppressed
 in generic hadron - hadron collisions.
  The reason is that, if there is a parton interchange  in the projectile/target or both,
 the large  logarithm is lost, which is due to the integration over transverse momenta. As a result such diagrams are not double collinear enhanced and do not contribute in the LLA (the authors of \cite{BR} call this type of diagrams the ladder cross talk). The physical
reason is that in order to get a large  logarithm from the integration over transverse momenta in the ladder the partons in the initial and final states must be at the same impact parameter.
While this occurs automatically for diagonal pairing, this generally does not happen for pairing of arbitrary partons. There is an additional small factor due to the longitudinal color delocalization in such exchange  as the color interchange creates a color dipole of length comparable to the nucleon size and hence carrying a significant excitation  energy \cite{BDFS3}.
\par The only way to avoid loosing transverse logarithm is to consider the $1\otimes 2$ processes.
The interference  for the $1\otimes 2$ processes
was  studied recently by Gaunt \cite{Gaunt},   In this case   two partons which interact with the deuteron are created in the split of a single parton of the projectile
nucleon. They are located at the same impact parameter. Hence  such interference diagrams contributing in the LLA (double collinearly enhanced). However the contribution
of this mechanism may become sizable only at very small x, near the black disk regime limit. Indeed, the contribution of  $1\otimes 2$ mechanism to the DPA is  is small in the discussed x-range. Thus the  interference contributions considered in \cite{Gaunt} are actually a small correction to already small
correction to DPA due to $1\otimes 2$ processes.

\par Indeed, it was showed in Sect.~2B that
the contribution of $1\otimes 2$ mechanism to DPA is $\sim 5 (10, 20)\%$
for  $Q^2=2 (10, 100) \mbox{GeV}^2$ respectively.
For our kinematics
typical $x$ are of the order 0.1 or larger.
In this case 
the interference is negligible
 relative to the full $1\otimes 2$
contribution\cite{Gaunt}.
 Hence the overall upper limit on the interference based on these considerations  is much smaller than  the shadowing effect which we calculated above.

\par 
At the same time
it  follows from analysis in \cite{Gaunt} that significant contribution of interference to $1\otimes 2$ can appear potentially, even for   symmetric kinematics
for very small $x$, since they are effectively defined by values of $x$ where the split occurs. Only then it can become comparable to shadowing.
This case needs further study, In
 particular a more detailed analysis of the ladder cross talk effect \cite{BR1}
 is desirable.

  \subsection{Color suppression  for a single interchange in the deuteron.}
\par We explained above that the interference contributions are small in the LLA. Here we shall show that there are additional suppression mechanisms that will
reduce interference further, even beyond the LLA.
  Let us now show that the interchange of two  partons between
  neutron
   and proton
  in the deuteron, in the case when no exchange occurs in the projectile proton,
  leads to the color suppression
 by a  factor $d_c$, where $d_c$ is the dimension of SU(3) irreducible representation
 to which the parton belongs.
 Such a suppression is a reflection of 
 the
  well known property
 of the suppression of nonplanar diagrams as compared to planar ones.
  For simplicity we shall consider the interaction of two partons
  of the deuteron with two partons of the projectile
   due to single  gluon  exchanges in t-channel.
 Indeed, consider for example
  the case of two baryons, $q^{i_1}q^{i_2}q^{i_3}..q^{i_{N_c}}$. Their wave functions in the color space are
  $\frac{1}{\sqrt{N_c!}}\epsilon^{i_1i_2...}q_{1i_1}q_{2i_2}q_{3i_3}...$ for the first nucleon and  $\frac{1}{\sqrt{N_c!}}\epsilon^{j_1j_2j_3..}q_{(N_c+1)j_1}q_{(N_c+2)j_2}...$
  for the second one.
  Consider the color factor from  the projectile nucleon. For simplicity assume that  two dijets originate from  quark-- quark
  scattering. The  color factor that we obtain from contracting the same
  quark in the amplitude and the conjugated amplitude
  is $\tr(t^at^{a'})\cdot \tr(t^bt^{b'})$,
  where we sum over final jet indices.
  The color factor from the projectile nucleon gives $\frac{1}{4}\delta^{aa'}\delta^{bb'}$.
  \par Consider now the factor originating from the deuteron
  block:
  \beq
  t^{a}_{si_1}q_{2i_2}....\epsilon^{i_1i_2..} t^{b}_{s_1p_1}q_{(N_c+2)q}q_{(N_c+3)r}...\epsilon^{pqr.....}.
  \label{tup1}
  \eeq

  The corresponding factor in the conjugated amplitude in the diagonal case is
\beq
    t^{a'}_{si'}q_{2j'}q_{3k'}...\epsilon^{i'j'k'...} t^{b'}_{s_1p'}q_{(N_c+2)q'}q_{(N_c+3)r'}...\epsilon^{p'q'r'....}.
    \label{tup2}
\eeq
Taking the product we obtain
\beq
\frac{1}{4N^2_c}\tr(t^at^{a'})\tr(t^bt^{b'})=\delta^{aa'}\delta^{bb'}\frac{1}{4N_c^2}.
\label{tup3}
\eeq
Combining color factors coming from the projectile and deuteron blocks we  finally obtain \beq
\frac{1}{4N_c^2}(N^2_c-1)^2.\label{tup4}\eeq

\par Consider now the interference term. In this case
 quarks "1" and "$N_c+1$"  are interchanged   between two nucleons in the conjugated amplitude, while
having the same initial state (here for simplicity we consider two nucleons consisting of  $N_c$ quarks with $N_c$ flavors). Hence the nucleon  wave functions in the conjugated amplitude are $\frac{1}{\sqrt{N_c!}}\epsilon^{ijk...}q_{(N_c+1)i}q_{(2)j}q_{(3)k}....$ for the first nucleon and  $\frac{1}{\sqrt{N_c!}}\epsilon^{pqr..}q_{1p}q_{(N_c+2)q}q_{(N_c+3)r}.....$ for the second one.
Then the color factor 
originating
from the deuteron block is
\beq
\frac{1}{N_c!} t^{a'}_{si'}q_{(N_c+2)j'}q_{(N_c+3)k'}\epsilon^{i'j'k'...} t^{b'}_{s_1p'}q_{2q'}q_{3r'}....\epsilon^{p'q'r'....}.
\eeq
 Calculating the product  we obtain
\beq
\frac{1}{N_c}(t^at^{b'})_{s's_1} \otimes \frac{1}{N_c}(t^bt^{a'})_{s_1's}.
\eeq
Taking the trace over indices of the final jets we obtain
\beq
\tr(t^at^{b'}t^{b}t^{a'}).
\eeq
Combining with the  color factor coming from the proton block we obtain
\beq
\frac{1}{N_c^2}\tr(t^at^bt^bt^a)=\frac{1}{N_c^2}c_F^2N_c=\frac{1}{N_c^2}(N^2_c-1)^2\frac{1}{4N_c},
\eeq
which is  $1/N_c$ smaller than in the diagonal case.
\par The same calculation can be done for  two dijets originating from the scattering off two gluons.
 For simplicity let us take the gluon part
of the first nucleon wave function as a color singlet $g_1^ag_2^a$, where gluon $g_1$ participates in the scattering process and the second one is a spectator,
while the second nucleon has wave function $g_3^ag_4^a$.Repeating the same calculation as for the quark case, we find that the factor
originating from the projectile nucleon  is
$N_c^2\delta^{aa'}\delta^{bb'}$. For the deuteron contribution for the diagonal case we get
$\tr(T^aT^{a'})tr(T^bT^{b'})=N_c^2\delta^{aa'}\delta^{bb'}$, where the matrices $T$ are generators in the adjoint representation. Combining
the factors coming from the projectile nucleon and the target deuteron  we obtain for diagonal case
\beq
N^4_c(N^2_c-1)^2.
\eeq
In the same way for the interference contribution  we obtain
$\tr(T^aT^bT^{b'}T^{a'})$, and after combining with the upper 
block
 of the diagram we get
\beq
c_V^2N_c^2(N_c^2-1)=N^4_c(N_c^2-1),
\eeq
which corresponds to  the $1/(N^2_c-1)$  suppression.
From these two examples it is clear that if we interchange the partons in the conjugated amplitude,  the interchanged parton being in irreducible representation
of $SU(3)$ with dimension $d_c$, we obtain the $1/d_c$ suppression. The similar arguments 
for spin variables
for
the
chiral states
 give suppression $1/d_s$,
 where $d_s$ is a number of spin states.
 So altogether we obtain a factor of 1/6 suppression for quark, and a factor of 1/16 suppression for gluon interference.

\subsection{Color suppression for a double interchange.}
\par Consider now double interference, in this case using the same arguments we see that if we interchange the partons both in the nuclear part (between two nucleons)
and in the upper part of the diagram, we get the product of two traces, i.e. for quark case we obtain
\beq
\tr(t^at^bt^{b'}t^{a'})\cdot \tr(t^{a'}t^{b'}t^{b}t^{a})\sim \frac{1}{2}\tr (t^{a'}t^{b'}t^{b}t^{a'}t^{b'}t^{b}).
\eeq
where the last equality is in the large $N_c$ limit. It is easy to see that in this limit the trace is $\sim N^2_c$, and thus the
double interchange 
 increases the color suppression to $1/d_c^2$, in the notations of the previous subsection. Note finally that such color suppressions
 were  included in the estimate of the interference in LLA
  discussed
  in subsection B.

\subsection{Accuracy of the two nucleon approximation for the deuteron.}
\par Finally, we assumed above that the deuteron in both initial and conjugated amplitudes consists of two nucleons.
Since the deuteron block for $\Delta=0$ corresponds to the intermediate state for the deuteron wave function  which is not a two - nucleon  state  we can use the information about the deuteron structure to estimate the probability of the non-nucleonic (exotic) component of the deuteron wave function, $P_{ex}$ as well.
The exotic components are  expected to have a small probability   since  the $NN\pi$ configurations are  suppressed by the chiral nature of the pion \cite{FS88}, while
 the lowest mass  two baryon intermediate state is    $\Delta \Delta$ , which has   a mass gap of $\sim 2(m_{\Delta} - 2m_N \sim \mbox{600 MeV}$ with the ground state. As a result one expects that   the probability of the non-nucleonic component in the deuteron is  $P_{ex} \le (1\div 2)\cdot  10^{-3}$\cite{FS88}. The experimental limit on the probability of the non-nucleonic components in the short-range correlations (SRC)  in nuclei coming from the Jlab and BNL experiments is $\sim 0.1$, for a review see \cite{SRC}. Since the structure of SRC  in the deuteron and heavier nuclei is found to be very similar  and the probability of SRC in the deuteron is $\approx .04$  the current data
 lead to  the  upper limit for the exotic
 admixture  $P_{ex} (D) < 4\cdot  10^{-3}$.
  Note here that a likely candidate for the dominant  exotic component for the deuteron wave function, the lightest baryon intermediate state -- $\Delta \Delta$  cannot be generated via
   interchange of two gluons.

Complementary way to look at the problem is to consider the singularities in the t-channel for the parton interchange - in the case of the two gluon interchange the closest singularity is presumably a gluonium state which has a mass $m_{gluonium} \sim $ 1.5 GeV and hence corresponds to exceedingly small inter nucleon distances  which occur in the deuteron with probability on the scale of $10^{-3}$. Note also that this argument does not include a small factor due to the requirement the both nucleons after interchange of partons remain nucleons rather than some excited states since typically  the color is  delocalized in such  exchanges at the distance scale of the order on the nucleon size.

\par
Overall we see that the interference mechanism contribution is negligible in the leading twist
 LLA approximation, unless we
 consider
 kinematics region close to the black
disk regime, where the interference effects may be significant, but this region is clearly beyond the scope of this paper. In addition, we have seen 
that
there are
additional suppression mechanisms, like color/spin suppression, overlap of momentum flows (subsection A) that likely diminish the interference mechanism in an independent way.
More studies are necessary for $x, Q^2$ range
 for near the black disk limit. Going beyond the LLA is also highly desirable both for $pp$ and $pA$ scattering.
The case of large A will be considered elsewhere.

\section{Conclusion.}
\par We  calculated the contributions of DPA and the  nucleon shadowing  to the four  jet MPI cross section in the proton - deuteron  collisions in the limit when
 one of the probes has small $x$ and another has $x,Q^2$ in the range where shadowing is small.
 We have
 demonstrated
 that shadowing increases with the decrease of $x$, and decreases rapidly with the increase of hard scale.
 For large $p_t$ of one of the probes corresponding to a typical   jet
 trigger in $pA$ collisions at the LHC  and small $p_t$ of the
 other probe we obtain correction  of the order of $30 \%$.
 This contribution is not reduced to the substitution  of the  deuteron pdf instead of nucleon pdf in the impulse approximation formula -- it  is twice as large as a such naive guess. There is a reduction by the factor
 $1/(1+N)$ that may be of order
 1/2, depending on kinematics,  due to a completely different mechanism of $1\otimes 2$ enhancement of the four jet cross section.
We also provided arguments for the dominance of the leading twist shadowing
one nucleon - two nucleon interference mechanism over the contribution due to the interchange   of partons  between two nucleons
in the kinematics discussed ($x_1\le 0.1,x_2\ge 0.1,Q_1^2$ few GeV$^2$).  In particular we demonstrated that in the LLA used in our analysis the interference diagrams are strongly suppressed.
 Further studies of interference beyond LLA and in different
 kinematic domains  are desirable.
  This is
especially true in the region of small $Q^2$ and $x$, in proximity to the black disk regime.
\par Our analysis  will serve as a starting point to a
 more complicated calculation of shadowing in the case of heavy nuclei for the similar kinematics. Further studies will be necessary for calculations of the shadowing in the kinematics when both $x's$ of the partons from the nucleus
 are small  and hence more complicated diagrams contribute to the
 the nuclear  shadowing.

\acknowledgements We thank CERN theory division for hospitality during the time this work has started, and Yu.~Dokshitzer, L.~Frankfurt,
D.~Treleani and U.~Wiedemann
for useful discussions.
 \appendix
\section{Correspondence with the  Glauber model of $pA$ scattering}
It is easy to see that the structure of the double scattering term is very close to that for the double scattering term for the total cross section of $pA$ scattering
in the Glauber model. This similarity  holds
for any nuclear wave functions as the two-body form factor which enters in both cases is the same. Since the relevant expressions for the heavy nucleus case were derived before in \cite{ST} it is convenient to check the correspondence taking
the limit of large A, and neglecting nucleon - nucleon correlations.

The ratio of the double and single scattering terms in the Glauber series for the total cross section of $ hA$ scattering:
\begin{equation}
\sigma_{tot}^{hA}= \int d^2b 2(1- \exp(-\sigma_{tot}T(b)/2) = \sigma_1 -\sigma_2 + \sigma_3 -  ...  ,
\eeq
is given by
\begin{equation}
\sigma_2/ \sigma_1= {1\over 4} \sigma_{tot} \int T^2(b)d^2b/A.
\end{equation}
This expression
 differs from the ratio of  the cross section of production of four jets in the interaction with two and one nucleons (Eqs. \ref{2},\, \ref{11})   by the  factor of ${1\over 4}$ and substitution $\sigma_{tot}\to \pi R^2_{\rm int}$. The factor of four could be understood on the basis of  the AGK cutting rules
 \cite{AGK}  which
 state that  the double cut diagram enters with the  extra factor of two as compared to the shadowing correction to the total cross section. An another factor of two  reflects combinatorics of emission of "pair one" from either  first or second nucleon.

Using this observation  it is straightforward to find the  expressions for the double interaction contribution if the expression for the shadowing for the total cross section is
known (including the effects of nucleon - nucleon correlations)

For example, in the case of the scattering off the deuteron
contribution of the diagram 2 to $G_2(x_1,x_2,  \vec{\Delta} )$ is
given by (For the discussion of proton - deuteron  four  jet production in the coordinate space representation see \cite{GT1,GT2,GT3}.)
\begin{equation}
{_2G}^{D}(x_1,x_2,\vec{\Delta})= 2 G_N(x_1,\vec{\Delta})G_N(x_2,\vec{\Delta})\cdot  S_D(\vec{\Delta}),
\label{G2}
\end{equation}
Here  $S_D(\vec{\Delta})$ is the standard deuteron form factor defined above (Eq.~\ref{11b}), which enters in the Glauber double scattering term .
Factor of two in Eq.~\ref{G2} is due to combinatorics (the  factor of A(A-1)). This is just the result obtained in section II  - Eq.~\ref{11}.

Similarly, one can obtain the  expressions for the triple MPIs matching the corresponding expressions  of ref.~\cite{ST}.


\begin{thebibliography}{99}
\bibitem{ST} M.~Strikman and D.~Treleani,
  Phys.\ Rev.\ Lett.\  {\bf 88}, 031801 (2002)
  [hep-ph/0111468].
\bibitem{BSW}B.~Blok, M.~Strikman and U.~A.~Wiedemann,
  Eur.\ Phys.\ J.\ C {\bf 73} (2013) 2433
  [arXiv:1210.1477 [hep-ph]].
\bibitem{DS1}D.~d'Enterria and A.~M.~Snigirev,
  Phys.\ Lett.\ B {\bf 718} (2013) 1395
  [arXiv:1211.0197 [hep-ph]].
\bibitem{DS2} D.~d'Enterria and A.~M.~Snigirev,
  Phys.\ Lett.\ B {\bf 727} (2013) 157
  [arXiv:1301.5845 [hep-ph]].
\bibitem{GT1} G.~Calucci, S.~Salvini and D.~Treleani,
  arXiv:1309.6201 [hep-ph].
\bibitem{GT2} D.~Treleani and G.~Calucci,
  Phys.\ Rev.\ D {\bf 86} (2012) 036003
  [arXiv:1204.6403 [hep-ph]].
\bibitem{GT3}
  G.~Calucci and D.~Treleani,
  ``Nucleon-deuteron collision as a probe of the partonic distributions,''
 In proceedings 40th International Symposium on Multiparticle Dynamics (ISMD 2010)
21-25 Sep 2010. Antwerp, Belgium,e-book http://inspirehep.net/record/981125 pp.319-324.
\bibitem{Alice} E.~Scapparone [on behalf of the ALICE Collaboration],
  arXiv:1310.7732 [hep-ex].
\bibitem{Alice1}B.~Abelev {\it et al.}  [ALICE Collaboration],
  Phys.\ Lett.\ B {\bf 719} (2013) 29
  [arXiv:1212.2001].
\bibitem{Atlas}G.~Aad {\it et al.}  [ATLAS Collaboration],
  Phys.\ Rev.\ Lett.\  {\bf 110} (2013) 182302
\bibitem{CMS}S.~Chatrchyan {\it et al.}  [CMS Collaboration],
  Phys.\ Lett.\ B {\bf 718} (2013) 795
  [arXiv:1210.5482 [nucl-ex]].

\bibitem{BDFS1}
  B.~Blok, Yu.~Dokshitzer, L.~Frankfurt and M.~Strikman,
  Phys.\ Rev.\  D {\bf 83}, 071501 (2011)
  [arXiv:1009.2714 [hep-ph]].
\bibitem{BDFS2} B.~Blok, Y.~.Dokshitzer, L.~Frankfurt and M.~Strikman,
   Eur.\ Phys.\ J.\ C {\bf 72} (2012) 1963  [arXiv:1106.5533 [hep-ph]].
   \bibitem{BDFS3} B.~Blok, Y.~.Dokshitzer, L.~Frankfurt and M.~Strikman,
arXiv:1206.5594 [hep-ph]
\bibitem{BDFS4} B.~Blok, Y.~.Dokshitzer, L.~Frankfurt and M.~Strikman,
arXiv:1306.3763 [hep-ph].
\bibitem{Guzey} L.~Frankfurt, V.~Guzey and M.~Strikman,
  Phys.\ Rept.\  {\bf 512} (2012) 255
  [arXiv:1106.2091 [hep-ph]].

  \bibitem{psi} V.~Guzey, E.~Kryshen, M.~Strikman and M.~Zhalov,
  Phys.\ Lett.\ B {\bf 726} (2013) 290
  [arXiv:1305.1724 [hep-ph]];
  V.~Guzey and M.~Zhalov,
  JHEP {\bf 1310} (2013) 207
  [arXiv:1307.4526 [hep-ph]].
\bibitem{FSW}
  L.~Frankfurt, M.~Strikman and C.~Weiss,
  Phys.\ Rev.\ D {\bf 69} (2004) 114010
  [hep-ph/0311231].
  \bibitem{FS99}
  L.~Frankfurt and M.~Strikman,
  Eur.\ Phys.\ J.\ A {\bf 5} (1999) 293
  [hep-ph/9812322].
  \bibitem{Gribov:1968jf}
  V.~N.~Gribov,
  Sov.\ Phys.\ JETP {\bf 29} (1969) 483
   [Zh.\ Eksp.\ Teor.\ Fiz.\  {\bf 56} (1969) 892].
   \bibitem{AGK}  V.~A.~Abramovsky, V.~N.~Gribov and O.~V.~Kancheli,
  Yad.\ Fiz.\  {\bf 18} (1973) 595
   [Sov.\ J.\ Nucl.\ Phys.\  {\bf 18} (1974) 308].
    \bibitem{HERA} S. Chekanov et al [Zeus Collaboration],Nucl. Phys., B816 (2009) 1.
  \bibitem{Paver:1984ux}
  N.~Paver and D.~Treleani,
  Z.\ Phys.\ C {\bf 28} (1985) 187.
\bibitem{mufti} M.\ Mekhfi, Phys. Rev. D{\bf 32}, 2371 (1985).
  \bibitem{LS}
   V.N.Kolybasov and M.S. Marinov,
   Sov. Phys.-Uspehi, 109(1973) 137.
   \bibitem{Lipatov} B. Ioffe, V. Fadin and L.Lipatov, Quantum Chromodynamics, University Press, Cambridge, 2010.
   \bibitem{FS88}L.~L.~Frankfurt and M.~I.~Strikman,
  Phys.\ Rept.\  {\bf 160}, (1988) 235.
  \bibitem{SRC}
  L.~Frankfurt, M.~Sargsian and M.~Strikman,
  Int.\ J.\ Mod.\ Phys.\ A {\bf 23} (2008) 2991
  [arXiv:0806.4412 [nucl-th]].
  \bibitem{BR}
  J.~Bartels and M.~G.~Ryskin,
  arXiv:1105.1638 [hep-ph].
  \bibitem{Gaunt}
  J.~R.~Gaunt,
  JHEP {\bf 1301} (2013) 042
  [arXiv:1207.0480 [hep-ph]]; talk at MPI@LHC 2013 meeting, Atwerpen, December 2013.
  \bibitem{BR1} J.~Bartels and M.~G.~Ryskin,
  Z.\ Phys.\ C {\bf 60} (1993) 751.
\end{thebibliography}
\end{document}